\documentstyle{cupconf}

\ifoldfss
\else
  \ifnfssone
    \newmathalphabet{\mathit}
      \addtoversion{normal}{\mathit}{cmr}{m}{it}
      \addtoversion{bold}{\mathit}{cmr}{bx}{it}
    \newmathalphabet{\mathcal}
      \addtoversion{normal}{\mathcal}{cmsy}{m}{n}
    \else
    \ifnfsstwo
    \fi
  \fi
\fi

\def\etal{{\it et~al.}}
\def \m{\ifmmode M_\odot\else M$_\odot$\fi}
\def \lta {\mathrel{\vcenter
     {\hbox{$<$}\nointerlineskip\hbox{$\sim$}}}}
\def \gta {\mathrel{\vcenter
     {\hbox{$>$}\nointerlineskip\hbox{$\sim$}}}}

\def\hexnumber#1{\ifcase#1 0\or1\or2\or3\or4\or5\or6\or7\or8\or9\or
 A\or B\or C\or D\or E\or F\fi }

\makeatletter
\ifx\CUP@mtlplain@loaded\undefined
\else
\fi
\makeatother
 \makeatletter
 \ifx\CUP@mtlplain@loaded\undefined
   \font\tenbmi=cmmib10 at 10pt
   \font\sevenbmi=cmmib10 at 7pt
   \font\fivebmi=cmmib10 at 5pt

   \newfam\bmifam
   \textfont\bmifam=\tenbmi
   \scriptfont\bmifam=\sevenbmi
   \scriptscriptfont\bmifam=\fivebmi
   
 \fi
 \makeatother
\ifnfsstwo

\fi
\ifnfssone

\fi
\ifoldfss

\fi

\mathchardef\varLambda="0103

\makeatletter
\ifx\CUP@mtlplain@loaded\undefined
\else
\fi
\makeatother
\makeatletter
\ifx\CUP@mtlplain@loaded\undefined
  \font\tenbms=cmbsy10
  \font\sevenbms=cmbsy10 at 7pt
  \font\fivebms=cmbsy10 at 5pt
  \newfam\bmsfam
  \textfont\bmsfam=\tenbms
  \scriptfont\bmsfam=\sevenbms
  \scriptscriptfont\bmsfam=\fivebms

  \edef\bsy@{\hexnumber\bmsfam}
  \mathchardef\bnabla="0\bsy@72
\fi
\makeatother
\def\etal{\mbox{\it et al.}}

\title[Black Hole X-ray Transients]{Black Hole X-ray Transients }

\author[J. Craig Wheeler]%
{J.\ns C\ls R\ls A\ls I\ls G\ns W\ls H\ls E\ls E\ls L\ls E\ls R\ls$^1$}

\affiliation{$^1$Department of Astronomy, University of Texas,
Austin, TX 78712, USA\\[\affilskip]}

\begin{document}
\ifnfssone
\else
  \ifnfsstwo
  \else
    \ifoldfss
      \let\mathcal\cal
      \let\mathrm\rm
      \let\mathsf\sf
    \fi
  \fi
\fi

\maketitle

\begin{abstract}

The observations and theory of the exciting new class of
galactic black hole X-ray transients is reviewed.  
Seven of these systems have measured
mass functions or mass estimates in excess of stable neutron stars,
making them excellent black hole candidates. Two of them have
revealed ``superluminal" radio jets.  Study of the hard and
soft radiation from these sources has given tight constraints 
on the physics of the viscosity of the accretion disk and
promises firm proof that these systems contain black holes.  This
will allow us to search for black holes of more moderate mass
and apply the knowledge of these systems to suspected supermassive
black holes in AGN's.  The most plausible mechanism 
for triggering the outburst of black hole candidate 
X-ray transients is the ionization thermal instability.
The disk instability models can give the deduced mass flow 
in quiescence, but not the X-ray spectrum.  Advection models that
can account for the quiescent X-ray spectrum are difficult to
match with the non-steady state, quiescent Keplerian disks. 
Self-irradiation of the disk in outburst may
not lead to X-ray reprocessing as the dominant source of optical light,
but may play a role in the ``reflare."  The hard power-law spectrum
and radio bursts may be non-thermal processes driven by the flow of
pair--rich plasma from the disk at early times and due to the formation
of a pair--rich plasma corona at late times.  The repeated outbursts
in systems like GRO J0422+32 suggest some sort of clock, 
but it is unlikely that it has anything to do with a simple X-ray 
heating of the companion star.     
These systems typically have low mass secondaries and their
evolutionary origin is still mysterious. 
\end{abstract}

\firstsection 
\section{Introduction}


In the past, Igor Novikov helped us to think about the 
inside of black holes and the methods to search for black
holes.  These subjects are today an active part of his legacy.  
In pursuing the astrophysics of
black holes it is useful to remind ourselves about the 
subject of our quest lest we become too
blas\'e or conclude the quest is finished.

Some black holes may arise from stars, but it is critical to remember 
that black holes are not ordinary
stars, even exotic ones such as white dwarfs or neutron stars.  
They may come from matter, but
they are not matter.  Black holes represent the ultimate construct 
of strong gravity, of curved
space time.  Even though black holes harbor their secrets within 
the shroud of the event horizon, they
are our conceptual link to the ultimate physics problem
of singularities, fundamental notions of
information as they evaporate, and to related concepts such 
as wormholes and closed time-like
curves.  It is thus of great importance to establish beyond any 
doubt that our Universe contains
black holes and then proceed to study them, probe them, learn from them.  
Does our Universe contain black holes?

How does one go about the search for black holes?  
There are perhaps 10$^{11}$ stars in the
Galaxy, around 10$^8$ neutron stars, and maybe a million 
black holes left from the deaths of
massive stars throughout the history of the Galaxy.  
The vast majority of these black holes will
be isolated and thus hidden from any search.  
Logic suggests that because stars tend to
congregate with like stars and black holes are thought to 
arise from massive stars that we
should restrict our search to bright, hot, massive stars 
that might orbit a previously deceased
companion, now a black hole.  Massive stars are, however,
rare. If the black hole is the  
product of a rare, massive star and it is in orbit 
around another intrinsically rare massive
star while that star lives its relatively brief life
then the likelihood of the combination is low.  The 
result is that there may be only
one such system in our Galaxy and we have found it:  
the venerable Cygnus X-1.  

It was with rather great surprise, then, that the last several years 
have taught us that the most common
form of binary black holes candidates in the Galaxy, perhaps 100 to 
1000 of them (Tanaka 1990), are black holes orbiting
very small, but intrinsically common, half-solar mass K dwarfs.  
These systems are proving to be
far more common than those like Cygnus X-1, but also to 
provide a fascinating laboratory to
study the astrophysics of black holes.

We are one hundred percent sure we have discovered neutron stars.  
Have we discovered black
holes?  The probability is $P=1-\epsilon$.  
Many may think that $\epsilon=0$ given the great
work of HST (Ford \etal\ 1994) and the very clever result based on 
OH maser emission by Miyoshi \etal\ (1995) showing that
active galaxies probably harbor massive black holes, as
long suspected.  In addition, at least 
six of the black hole transients
to be discussed below have mass functions and 
hence minimum masses in excess of 3 \m, the
neutron star limit.  One might, however, argue that this 
evidence is still stubbornly circumstantial
given that we simply know these objects have high gravity, 
large mass and no optical stellar
counterpart.  That might leave $\epsilon$ finite, perhaps a few percent.  
What do we need to provide
incontrovertible {\it proof} that we have discovered black holes?  
The black hole X-ray
transients may help to yield that proof.

Section 2 will give an overview of the observations.  
Some discussion of the relevant disk instability that
probably triggers the outbursts and issues of self-irradiation of
the disk and reprocessing is given in \S\S 3 and 4.  
Section 5 describes the important lessons we have learned
from the exponential decline about the physical nature
of the disk viscosity.  The curious ``reflare" phenomenon
is discussed in \S 6 and various aspects of the new
advection solutions are given \S 7.  
The quasar-like hard power law is considered in 
\S 8.  Section 9 describes some of the problems of 
progenitor evolution and related issues.  A summary and
outline of key issues is given in \S 10.

\section {Observations}

There are by now upward of 24 binary X-ray sources that have 
been suggested as black hole
candidates (Tanaka \& Lewin 1995; Chen \etal\ 1996).  
Six of these have measured mass functions that
make them excellent candidates on that basis alone:  
AO 620-00 (Nova Mon 1975), $M_x>3$ \m\ (McClintock \& Remillard 1986); 
H 1705-250 (Nova Oph 1977), $M_x>4$ \m\ (Remillard \etal\ 1996);
GS 2000+25 (Nova Vul 1988), $M_x>5$ \m\ (Filippenko, Matheson, and Ho 1995);
GS 2023+338 (V404 Cyg), $M_x>6$ \m\ (Casares \& Charles 1994;
Filippenko, Matheson, \& Barth 1995); 
GS 1124-683 = GRS  1121-68 (Nova Muscae 1991), $M_x>3$ \m\ (Remillard,
McClintock, \& Bailyn 1992); 
GRO J1655-40 (Nova Sco 1994), $M_x>3$ \m\ (Bailyn \etal\ 1995).
Other systems are placed in this category 
of black hole candidates because
they share morphological similarities to these best candidates, 
for instance in terms of their
light curves or spectra or have other confirming orbital data.  
Among the more prominent of these are LMC X-1, LMC X-3, GX 339-4,
1E1740.7, GRO J0422+32 (Nova Persei 1992), GRS 1716-249 (Nova Ophiuchi 1993), 
GRS 1009-45 (Nova Vel 1993) and GRS 1915+105.  
Note that after a hiatus between 
the discovery of the prototype
A0620-00 and that of GS 2000+25 in 1988, candidate 
systems have been observed at the rate of about
one per year.  Since 1992 this has been in great part 
due to the effectiveness of the
BATSE experiment on CGRO (Fishman \etal\ 1989; Harmon \etal\ 1994).  
This discovery distribution is a powerful argument
for the maintenance of all-sky monitors
in a variety of wavelengths.

Some of the black hole transients are known to 
have repeated, A0620-00 in 1916 and 1975, V404
Cyg in 1938, 1989, and perhaps in 1956,
and all of them are suspected 
to have done so on time scales of one to a few decades.
Some repeat in about a year, but with properties
that may be different than the major outbursts. 
Nova Sco which was first discovered in 1994
is undergoing another outburst at this writing
(May 1996).

Many of the outbursts are characteristically marked by a very rapid
rise on the time scale of about a day   
and then a nearly exponential decline in soft X-rays with 
an e-folding time of 30--40 days. 
This behavior was clearly displayed by A0620-00, GS200+25 
and Nova Muscae 1991.  This form of the light curve 
is not universal, however.  V404 Cyg had a very 
irregular light curve near maximum.  Other
potential candidates have shown other types of behavior, 
plateaus, triangles, etc (Chen \etal\ 1996).  A caveat here is that
the fast rise and 30--40 day decay is a signature of 
the soft X-rays.  The hard X-rays, with a
manifestly different physical origin, often display burst behavior, 
but with a rather different
morphology than the soft X-rays.  
If one only monitored the hard X-rays, 
as is the recent circumstance with BATSE but
without the regular coverage of Ginga, then the picture 
one gets is rather different.  It is thus
difficult to know whether or not the various light 
curve morphologies BATSE has observed belie a very
different underlying structure of the object, 
accretion disk, outflows, coronae, etc.
or whether or not one is seeing the tail, not the dog, because
of energy response limited to $\gta$ 20 keV.

Another common feature of the light 
curves of the black hole candidates is a
secondary brightening sometimes called a ``reflare" at 
60--80 days after maximum.  This feature is not universal,
but has been observed in all the sources with the canonical
fast rise and exponential decay.   Such a
feature has never been observed in a neutron star soft 
X-ray transient with fast rise an exponential decline (Aql X-1 or Cen X-4), 
so there is some suspicion that it, too, is a signature of a black hole.  
V404 Cyg did not show this clear ``reflare" and 
GRO 0422+32 showed only a bare
flattening at $\sim$50 days.  On the other hand the ``reflare" is 
a feature of the soft X-rays,
presumably arising from the geometrically thin, 
optically thick inner accretion disk.  Neither
V404 Cyg nor GRO J0422+32 showed this soft X-ray component, 
but only the hard component which
has some other physical origin.
Some of the most interesting recent sources, the 
``super-luminal" sources GRS 1915+105 and GRO J1655-40
to be discussed below, showed irregular light curves in the 
BATSE band, with little correlating information. 

Where observations have been sufficient, many systems have 
shown a ``second maximum" in flux
about 200 days after the primary outburst, among 
them A0620-00, Nova Muscae, and GRO J0422+32. 
This feature can be observed in both soft and hard X-rays, 
but the soft component is probably the low
energy extension of the hard power law component and not 
at all from the accretion disk.

Two systems, GRO J0422+32 (Callanan \etal\ 1995)
and Nova Vela 1993 (Bailyn \& Orosz 1995; Della Valle, Benetti
\& Wheeler 1996), have shown yet more subsequent outbursts.  These
bursts have been detected primarily in the optical 
with very irregular coverage in other bands,
soft and hard X-ray, radio.  It is thus difficult 
to determine their true nature and the
connection to previous flux peaks.

Many of the black hole candidates show a characteristic 
two-component spectral structure.  One component
is a power-law that extends to high energies $>$20 keV perhaps 
$>$400 keV.  As mentioned above, this power-law 
presumably also extends to lower energies.  
Rising above this power law at
low energies, a few keV, is an extra ``ultra-soft" 
X-ray component that is thought to arise in the
accretion disk.  These two components do not vary in lock step.  
As the soft component declines
more or less monotonically in the characteristic exponential decay, 
the hard component can vary
non-monotonically.  In some phases the behavior is 
complementary with the hard rising as the
soft declines, reminiscent of the different ``$\gamma$-states" 
of Cyg X-1 (Ling \etal; Crary \etal\ 1996).  In particular it is
important to note that while (one component of) the soft 
X-ray flux varies with the presumed
mass flow rate in the inner disk, the power law component 
does not behave so simply.  Its power
may ultimately derive from the release of gravitational energy, 
but there remains at least some
possibility that some of its energy derives from the black hole, 
perhaps in a Blandford-Znajek process (Blandford \& Znajek 1977) .

As noted above there are exceptions to this 
characterization of the spectral behavior.  The
``ultra-soft" component was definitely absent in V404 Cyg 
and GRO J0422+32 and Nova Oph 1993.  It is not clear
whether this could be some geometrical effect obscuring 
emission from the disk or a true absence
of the inner soft X-ray emitting accretion disk.  
Tanaka (1990) argues that the large amplitude noise of the
emission from V404 Cyg near maximum, a characteristic of 
the power law that is suppressed when
the soft X-rays dominate, is evidence that the inner 
disk component is absent.  On the other hand
GRO J0422+32 did not show the great variability of V404 Cyg.
Nova Oph had a strange square light curve (Harmon \etal\ 1994),
and it is not clear how it fits into the general scheme.

The short-term variability is also of great interest in these systems.  
Nova Muscae showed QPO's
at 3, 5, and 8 Hz (Miyamoto \etal\ 1993) 
and GRO J0422+32 at 0.04 and 0.1 Hz (Kouveliotou \etal\ 1993).  
The physics behind this variation is not
yet understood, but it is clear that black hole 
sources and neutron star sources can have rather
similar behavior and so the phenomena can not be restricted 
to any disk/magnetosphere interaction.
There is some suggestion that this sort of time variability
should be used to help classify the physical state of 
the system (van der Klis 1994, 1996).

Radio emission is also a ubiquitous feature of these 
systems where again no obvious parallel has
been observed in the neutron star analog transient systems.  
The radio emission tends to follow
the primary outburst, but most sources 
are not spatially resolved or well sampled temporally, so one does 
not know exactly where or when the radio was
generated.  This emission has been interpreted as an 
expanding synchroton bubble (Han and Hjellming 1992), 
but it may all be better interpreted as a jet (Hjellming 1996). 
The relativistic particles might be created at some ``working surface" 
that is somehow energized by power
from the underlying system, but if jets are the rule then
the fast particles almost certainly arise in the disk.  The manner in which
they are created is still not clear (\S8).  

More dramatically, two sources, GRS 1915+105 (Mirabel \& Rodriguez 1995)
and Nova Sco 1994 (Harmon \etal\ 1995; Hjellming \& Rupen 1995; 
Tingay \etal\ 1995) have shown well-collimated ``super-luminal" 
outflow. If the notion that we are dealing with black holes that emit
power-law hard radiation and radio synchrotron radiation 
were not already sufficient, the
discovery of these superluminal sources has completed the 
``mini-quasar" analogy.  The X-ray
light curves of both of these objects are somewhat 
unorthodox, but a well sampled 
soft X-ray light curve is lacking.
GRS 1915+105 showed a long slow rise and then a outburst in the
radio.  It has a classic symmetric double-sided structure.  Nova Sco
had a rapid rise, but an irregular structure in which the
radio bursts seems to correlate with a decline in the hard
X-ray.  It had an irregular hard X-ray light curve that
bears no simple relation to the fast rise and exponential
decline format.  The radio structure is more irregular,
with asymmetric blob structure.

Nova Muscae was well-sampled by Ginga and Sigma/GRANAT
which detected both the ultra-soft flux and sufficiently
high energy photons to sample the separate behavior of the power law tail.  
Miyamoto \etal\ (1993) have
provided a decomposition of the light curve that is both rich 
in import for the physics of this
and related systems, but also illustrative of the 
need for good sampling in both soft and hard
X-rays as well as other bands, e.g. optical and radio.

The total power of Nova Muscae in all Ginga bands rose to a peak 
of 10$^{38}$ erg s$^{-1}$ in about a day and
then declined exponentially following the soft X-rays.  
The power went up at the ``reflare" when
the power law component was at a minimum, clearly 
establishing that the ``reflare" is a phenomenon of the
soft flux from the accretion disk. The ``reflare"
is also observed in optical and UV, so either the outer disk is
involved or the mechanism of the inner disk can radiate in the optical.
The total power went through a local minimum around day 170 
and then another local maximum around day 200.
This ``second maximum" falls approximately on the extrapolation of
the exponential.  This may imply that the minimum is a decrease
in flow rate or efficiency, rather than the 200 day maximum being an
enhancement of either.  On the other hand, in A0620-00 the 
``second maximum" at 200 days has an amplitude considerably in 
excess of an extrapolation of the exponential, so it may
not be possible to draw general conclusions from this behavior.

Miyamoto \etal\ (1993) deconvolve the total power into 
the power law component, including an
extrapolation to soft X-rays, and the component ascribed 
to the disk, the excess in soft X-rays
above the power-law extrapolation.
The soft X-ray flux ascribed to the accretion disk 
rose at the ``reflare" as remarked above.  In A0620-00 there is
some suggestion that the soft X-ray flux declined with a steeper
slope after the ``reflare."  The data might also be compatible with
an increase of the soft X-ray flux to a higher level 
at the time of the ``reflare" with the same decline rate immediately
after the ``reflare" followed
by a drop in flux to to the original trend at about 100 days.    
For Nova Muscae the soft X-ray light curve
shows a rise at the ``reflare" followed by a decay with
a somewhat shallower slope before the decline at about 140 days
that precedes the ``second maximum" at 200 days.  GS2000+25 might show
a similar behavior, but the sparse data is open to considerable 
interpretation.  In the optical,
A0620-00 showed a flatter slope throughout the exponential
decline with an e-fold time of about 68 days,
much shallower than the 30 days of the X-ray.  The optical ``reflare"
was basically contemporaneous with the soft X-ray ``reflare," but
the flux returned to the same trend as before the event within
about 10 days.  There was an unfortunate paucity of optical
data on Nova Muscae.  The origin of this
``reflare" and the subsequent behavior of the light curve 
are not well understood, but may have something to do with
the self-irradiation of the disk as will be discussed below (\S4).  

In Nova Muscae the disk flux 
began to plummet about 140 days after peak and faded 
to near the threshold of detectability by 170
days, corresponding to the minimum in the total power curve, 
a factor of over 100 in 30 days.
There is no apparent contribution from the accretion disk 
to the ``second maximum" at 200 days. 
This shows that this local maximum may be entirely 
ascribed to the power-law component and not to the
disk, just the opposite of the ``reflare."

By deconvolving the spectrum into a disk and power law components 
the parameters that describe
the disk model can be constrained.  In particular the inner edge 
of the disk seems to be
remarkably constant in a number of sources 
(GS2000+25; Nova Muscae; LMC X-3) when the soft
X-ray luminosity and hence presumably the mass 
flow rate through the inner disk varies by factors
of 10--100.  There is only one obvious characteristic 
radius of the inner disk that should remain
constant independent of \.M and that is the last 
stable circular orbit that depends only on the
mass of the black hole.  Although somewhat uncertain, 
the absolute value of the derived inner
radius is also consistent with the last stable orbit 
of a black hole of a few \m (Miyamoto \etal\ 1993;
Ebisawa \etal\ 1994).  These
observations are remarkably consistent with what one 
expects from a black hole and inconsistent
with any kind of neutron star magnetosphere where the 
radius should adjust to the mass flow.

The hard component of the first outburst ascribed to the power-law
source by Miyamoto \etal\ (1993) peaked before the 
soft flux in Nova Muscae and then declined rapidly, much faster
than the disk flux.  It went through a local minimum about
five days after the power law peak, at about the time of
the peak of the disk component as deconvolved by Miyamoto
\etal\ and then rose again to begin a somewhat more slow decay.
As the power law began this new trend the transient feature
at 480 keV which has been ascribed to red-shifted annihilation
radiation was detected (Sunyaev \etal\ 1992; Goldwurm \etal\ 1993).  
The power law flux began a more rapid decline again about
30 days after the peak and went through another sharp 
local minimum at about 40 days after peak.  One wonders whether
there was another annihilation event at that time, but 
there were no appropriate observations.  The power law
component went through a minimum from 70 to 100 days after
peak and then began a rise to dominate the flux at the
``second maximum" at 200 days.  

Although the spectral index in the first
peak and the ``second maximum" were similar, 
it is possible that the physics of these
two events were rather different. The first was associated
with a radio burst and the apparent annihilation event.
There are reasons to think that this event was associated
with some form of strong dynamic outflow (Moscoso \& Wheeler 1994; \S8).
The ``second maximum" may be more similar to the classic coronae
discussed in the literature which are assumed to be in
hydrostatic equilibrium (Liang 1990).  Either of these configurations could
interdict the inner portion of the disk, thereby preventing
it from reaching to small radii where soft X-ray ($\sim$ keV)
can be radiated efficiently.  This may be why the hard flux
precedes the soft flux on the rise while the disk is ``filling in"
the inner region from which the outflow emerges and why
the disk flux plummets before the ``second maximum" as the ``corona"
swells to fill up the region previously occupied by the inner disk.
Unfortunately, the observations of Nova Muscae
were not sufficient to determine
whether the radius of the inner edge of the disk increased
as the ``disk" component declined at 140 days (Ebisawa 1996).

\section{Disk Instability}

The most plausible origin of the outburst in these systems 
is thermal instability associated with
the ionization of hydrogen and helium which is also 
thought to be the fundamental mechanism of
dwarf novae where the central star is a white dwarf 
(Meyer \& Meyer-Hofmeister 1981; Cannizzo, Ghosh, \& Wheeler 1982; Smak 1982; 
Faulkner, Lin, \& Papaliozou 1983;  Mineshige \& Osaki 1983; 
for a recent review see Cannizzo 1993a).  This mechanism arises as
matter accumulates in the disk and becomes denser and warmer.  
Hydrogen begins to ionize and
this increases the opacity.  In order to maintain 
thermal equilibrium between viscous heating
and radiative losses the density must decrease to 
maintain the optical depth in the face of the
rising opacity.  This leads to the characteristic 
``S-shaped" curve in the plane of surface
density and temperature.  The surface density of the disk 
cannot decrease, however, since matter is
accumulating.  Instead the disk structure is driven out of
thermal equilibrium.  It heats until it attains a hot state where H is
fully ionized and thermal equilibrium can be restored.  
The hot state has higher viscosity and
this leads to a mass flow in excess of the transfer 
rate from the companion.  The surface
density and temperature decline until a critical density 
is reached where hydrogen begins to
recombine.  At this point a cooling instability ensues 
returning the disk to a cold state.

This cycle is imposed whenever the transfer rate from 
the companion would require a portion of
the disk to be in the region of thermal instability.  
It depends quantitatively, but not
qualitatively, on the nature of the viscosity.  
The instability in geometrically thin Keplerian
accretion disks generates heating waves that 
can propagate throughout the disk raising it to the
bright, hot state that represents the 
outburst.  This outburst can start in the outermost
disk if the time scale for mass accumulation there is shorter
than the viscous time scale.  This depends on the mass transfer rate 
and the viscosity prescription in models.  In other circumstances,
the outburst can start in the middle or inner portions of the disk.
After the outburst, the surface density profile becomes more 
concentrated at smaller radii. This leads to a decrease in the 
surface density at large radii and the subsequent cooling instability
and cooling wave is automatically started near the outer
edge of the disk. The
subsequent cycles of heating and cooling to quiescence account
for the basic observations of dwarf novae (Cannizzo 1993b) and 
candidate black hole transients (Mineshige and Wheeler 1989;
Cannizzo 1993b, Cannizzo 1994, Cannizzo, Chen, \& Livio 1995).

It is important to note that in the steady state the emission from
the disk surface does not depend on the viscosity prescription
(Shakura \& Sunyaev 1973), but
only on the local mass flow rate \.M in the disk (aside from
some boundary condition restrictions at the edges).  
In the time dependent case, the flux does depend on the viscosity
prescription.  This is a two-edged sword.  It means the predictions
of the models depend on the uncertain physics of the angular
momentum transport, but it also means that observations
coupled with models with particular viscosity prescriptions
can constrain the physical nature of the viscosity.  The success
of this approach will be outlined below (\S5).  In general, the 
transfer rate into the disk will not be equal to the mass flow
rate through any radius in the disk and in particular not to
the mass flow rate through the inner disk that determines the
X-ray luminosity in the black hole models.  One must avoid
analyzing time-dependent situations with convenient, but
erroneous, steady-state assumptions.

There are various tests of the disk instability in
the context of the black hole transients.  The models naturally
gives a rapid rise and slower decline in the soft X-ray and
optical.  They can give an exponential decay as observed in
some sources (Cannizzo, Chen, and Livio 1995; \S 5).
The mass transfer from the companion in A0620-00
greatly exceeds that attributed to the soft X-rays from
the inner disk (Marsh, Robinson \& Wood 1994; McClintock, Horne,
\& Remillard 1995).  This shows that the disks are not in steady state in
quiescence, a principle prediction of the disk instability models.     
There has been difficulty in detecting the cooling wave that should
attend the decline (Cheng \etal\ 1992), but this may be because 
coronal heating yields a different color temperature than
the effective temperature predicted by basic models.

\section{Disk Irradiation}

The question of the effect of disk self-irradiation is an 
important one for any source that emits intense X-rays, that
is any neutron star or black hole system.  In cases like
Cygnus X-1 irradiation from the companion may also play a role,
but this has not yet been quantitatively investigated. 
Irradiation from the companion to the disk is not important in the systems
with small mass companions, but irradiation of the companion
stars by the disk may be.

        Van Paradijs and McClintock (1994, 1995) argue that the optical
luminosity in black hole transients may be, in analogy to
low mass X-ray binaries, dominated by X-ray reprocessing in
the disk.  Van Paradijs (1996) has extended this argument to investigate
whether the disk irradiation can alter the critical transfer rate
below which the disk instability sets in. 
If the disk is heated by irradiation it can be maintained
in a stable, ionized state even if the mass transfer rate
alone is insufficient to keep the disk ionized.
     On the other hand, Cannizzo (1994) has argued 
that the irradiation in the disk is negligible based on
his study of the viscosity parameter $\alpha$ and its relation to the
decay time scales in dwarf novae and X-ray novae.       
Huang and Wheeler (1989) and Mineshige and Wheeler (1989) noted
that disk instability models of black hole sources
with no irradiation give ample optical luminosity.  

	A first step toward the understanding of irradiation in
the context of time-dependent, irradiated disk models has
been taken by Kim, Wheeler, and Mineshige (1996a,b).  They
adopt two simple models of the irradiation: direct irradiation from the
innermost hot disk and irradiation as might be reflected by a corona or disk
atmosphere or chromosphere above the disk.
     The X-ray luminosity of the irradiation is given by
$$
   L_X(t)={\epsilon}{\dot M}_{in}c^2,
                                                \eqno(1)
$$
where the efficiency $\epsilon$ =0.057, and
${\rm {\dot M}_{in}}$ is the mass accretion rate at the inner
edge of the disk $R_{in}$, taken to be 3$r_g$ for a Schwarzschild black hole.
     For the indirect irradiation the assumption is made that the
luminosity of Eq. 1 effectively arises from a point at the center of
the disk (R = 0).
     The indirect irradiation flux is then given by
$$
     F_i(t) = C_X {\Bigl({{L_X(t)}\over {4{\pi}R^2}}}\Bigr),  \eqno(2)
$$
where $C_X$ is a constant.
     More sophisticated models incorporating radiative transfer would give
$C_X$ = $C_X(R, t)$ (see discussion in 
Tuchman, Mineshige \& Wheeler 1990), but such
time-dependent radiative transfer is too difficult at this time.
     For the indirect irradiation, the prescription of Fukue (1992) is 
followed.  This model assumes that the irradiation from a hot, geometrically
thin annulus near the inside of the disk
can be approximated by that from an infinitesimally
thin, filled, uniformly radiating surface centered on the black hole.
For regions in the disk at distances much larger than
the radius of the annulus, this leads to the following expression:
$$
     F_d(t)=(1-A) {\Bigl({{L_X(t)}\over {2{\pi}R}}\Bigr)}
              {d\over dR}
              {\Bigl({{H(t)}\over R}\Bigr)}^2.                \eqno(3)
$$
     For simplicity, we take the X-ray albedo, $A$= 0.5.
     The specific choice of albedo does not appreciably affect the disk
structure since shadowing suppresses the direct irradiation (see below).
     Eq. (3) shows that the direct irradiation is a function of the gradient of
the disk height and hence very sensitive to small variation in the disk profile.
     Once ${\dot M}_{in}$ and hence $F_i(t)$ and $F_d(t)$ are determined, the
irradiated flux is added to the flux generated internally in the disk by
viscous heating through an implicit numerical method, and hence other disk
physical variables are corrected in each time step
(see Tuchman {\it et al.} 1990 and Mineshige, Tuchman \& Wheeler 1990 
for details).
     In this formulation, the irradiation fluxes are
parametrized by two constants,
$C_X$ and $A$, which are, by assumption, independent of radius and time.   

	Kim \etal\ (1996a,b) show that even relatively mild irradiation 
from the inner disk can affect the overall disk evolution.
     Irradiation enhances depletion of mass from the disk during outburst, 
which results in a longer quiescent interval.  
	It can also affect the slope of the declining light curve.  
	More specifically, irradiation can affect the limit cycle 
by causing portions of the disk to linger in the intermediate 
temperature metastable ``stagnation state."
     Although the disks are in nearly steady state near maximum, the
time dependent models show that this is not strictly so and this 
has important implications for the irradiation.  There tends to be 
a local maximum in H/R at intermediate radii ($\sim 10^8$ cm)
that causes the outer disk to fall in the shadow of the inner disk,
even at the time of maximum X-ray flux.  This strongly limits the 
effect of the direct irradiation.
Direct irradiation does not play an important
role in the outer disk during and after the maximum of the outburst,
but can affect the middle of the disk during the outburst.
     Since it is by assumption not influenced by shadowing, 
     the indirect irradiation affects the overall disk evolution,
including the outer disk, throughout its evolution. The indirect
irradiation is, unfortunately, very uncertain because it depends
on the X-ray scattering in coronae of unknown physical nature
which Kim \etal\ have not considered.   

Kim \etal\ have not explored the regime of strong indirect irradiation
(large $C_X$) in any detail on the grounds that the light curve
of the non-irradiated or only mildly irradiated models match the
observations reasonably well.  Such strong irradiation 
is likely to destroy this agreement, although this is a regime that
should be explored more thoroughly.
    It is true that strong irradiation
can heat the outer disk and stabilize disks that would otherwise
be unstable to the disk instability for a given disk size and 
mass transfer rate as argued by van Paradijs (1966).  
Van Paradijs adopts a steady-state model for direct irradiation and hence his
criterion for stabilizing the disks by irradiation 
may not apply to the transient sources. 
The time-dependence tends to yield disks with non-monotonic
height profiles and hence shadowing of the outer disk by the
inner and middle disk even in outburst when the disk is 
approximately in steady state.  This suggests that
transient sources should still follow the criterion given by, e.g.,
Smak (1983) neglecting irradiation.  Aql X-1 represents an interesting
case in point since it lies very close to the irradiated instability
line determined by van Paradijs.  One might argue that as a transient
it should be governed by Smak's criterion and be well below
the minimum transfer rate for stability.  Aql X-1 has a variety of
interesting properties that make it difficult to straightforwardly
interpret it in terms of a disk instability (Tanaka and Lewin 1995),
but it is worthy of closer examination in this respect.

	Further study is necessary to understand the correlation of the optical
activity to X-ray and radio observations.
	The irradiated models presented by Kim \etal\ 
are based on parameters chosen to reproduce
the basic behavior of the optical light curve of A0620-00.  They
show that the disk instability is initiated in the outer portions of
the disk.  This means that the optical should begin to rise
before any activity in harder bands.  Mass flow is, however,
enhanced in the inner disk before optical maximum, so soft
and hard X-rays might well precede optical maximum.  
These models give inner disk temperatures of several million degrees
near maximum, but this is not hot enough to provide  sufficient
soft X-rays.  
	Other choices of disk parameters might give larger
maximum flow rates and hence higher disk temperatures.

\section{The Significance of the Exponential Decay}

One of the interesting features of the black hole X-ray
novae is the tendency to show an exponential decline.
Simple models in which one quickly reduces the transfer rate
to a hot disk with with constant viscosity parameter
$\alpha$ generate geometrically
declining, not exponential, light curves.  Even models in
which the decline is driven by the cooling wave of the
disk instability tend to have geometrically declining light
curves with constant $\alpha$.  Mineshige, Yamasaki, \& Ishizaka (1993) 
have argued that
to produce an exponential decline, the angular momentum of the
inner disk must be removed at a rate proportional to the
angular momentum.  They note that this tends to be the behavior
of disk instability models with $\alpha=\alpha_{0}(H/R)^{n}$
with n $\sim$ 1 - 2.  Cannizzo (1994) has also addressed this argument
by noting that both dwarf novae and the black hole transients
have exponential declines.  Cannizzo concluded that to
reproduce the exponential one needs $\alpha\propto R^{\epsilon}$,
with $\epsilon\sim~0.3 - 0.4$,
which is consistent with Mineshige \etal\  Cannizzo carried
the argument one step further, however, by making the case that
the precise value of $\epsilon$ that leads to exponential
decline is itself a function of other parameters of the problem
such as the transfer rate and inner disk radius.  From this
he concluded that exponential decline requires some form of
feedback to operate in the disk to give just this behavior.
This may hint that the angular momentum transport process
is non-local, as the theories where internal waves play a
critical role imply (Vishniac \& Diamond 1993).

These arguments have been extended significantly by Cannizzo,
Chen, and Livio (1995).  
Cannizzo {\it et al.} have shown that finely zoned (non-irradiated)
models give a cooling front width w $\propto$ $\sqrt{HR}$
and that the light curve in decline can be exponential only
if there are certain conditions on $\alpha$.  Cannizzo \etal\
express the speed of the cooling front as,
$$
     {\rm V}_f~=~{2\over 3}\alpha_0 \left( {H\over R} \right)^{n+1}
      {c_s\over w}R,
                                \eqno(4)
$$
where $H$~=~$c_s{\Omega}$ is the vertical scale height at radius R,
c$_s$ is the sound speed, $\Omega$ the Keplerian angular velocity,
w is the width of the cooling front and the viscosity
parameter is given by $\alpha$ = $\alpha_0(H/R)^n$.
They argue that that the relevant sound speed is that
dictated by the temperature of recombination in the matter just
preceding the cooling front which is essentially constant,
giving $(H/R)$ $\propto$ $R^{1/2}$.
Assuming the criterion for exponential decay to be ${\rm V}_f$ $\propto$ R
the relation w$\propto\sqrt{HR}$ then
yields the observed exponential decay only
when n~=~3/2.  

Vishniac and Wheeler (1996) have amplified this
issue by showing that the cooling front speed is set by
the rarefaction wave (in surface density $\Sigma$) 
that precedes the cooling front
and that, in turn, determines the cooling front width, rather than
the other way around.
They show that the condition ${\rm V}_f$ $\propto$ R does lead
to an exponential decline in the mass in the inner hot region
and an exponential decline in \.M, but that the speed of the front
depends on opacities.  For Kramer's opacity, the value of n
that leads to an exponential decline is closer to n = 1.65 rather
than 1.5.  This preferred value
of n is consistent
with the theory of angular momentum transport by an internal wave-generated
dynamo driven by tidal instabilities at the outer edge of the disk
(\cite{VJD90}, \cite{VD92}).  
Vishniac and Wheeler argue that the 
cooling wave propagation depends on the
viscous flow in the hot state and is nearly independent
of the actual cooling process and of the
state of the disk in the cool, quiescent material
that accumulates in the wake of the inward-propagating
cooling wave.  The speed of the cooling waves is determined by the rarefaction
wave that precedes them and is approximately $\alpha_F c_{F} (H/R)^q$,
where $\alpha_F$ is the dimensionless viscosity, $c_{F}$ is the sound speed,
and all quantities are evaluated at the cooling front.  The
scaling exponent $q$ lies in the interval $[0,1]$, depending on the
slope of the $(T,\Sigma)$ relation in the hot state; 
however, $q$ will be
close to $1/2$ for most models of the disk hot state. 
These results are insensitive to the structure of the
disk outside of the radius where rapid cooling sets in.  
Vishniac and Wheeler conclude that
the exponential luminosity decay of cooling disks is probably compatible
with the wave-driven dynamo model.  It is not compatible with models with
separate, constant values of $\alpha$ for the hot and cold states.

There are several conclusions to be drawn from this work
of Cannizzo \etal\ and Vishniac and Wheeler.
First, the whole analysis is predicated on the
assumption that a cooling wave exists in the decline of the
light curve of transient black hole candidates and related
systems.  While the evidence is indirect, one can thus regard
the exponential decline as a strong argument that a cooling wave
is the fundamental mechanism of the decline of these
transients. Since the cooling wave is one of the principal aspects
of the disk instability, the exponential decay adds to the evidence that
the accretion disk ionization instability is the
underlying physical cause of the transient outburst phenomenon.

Another important general conclusion is that the study of accretion
disks with phenomenological models for the angular momentum 
transport have provided crucial constraints on the nature of the
viscosity.  The quantitative and even qualitative behavior of the black
hole models depends on the prescription for $\alpha$.
In the case of a double valued, but radially constant
prescription, the outburst will tend to occur in the
inner disk, yielding a somewhat slower rise phase and more
symmetric outbursts.  A prescription in which
$\alpha=\alpha_{0}(H/R)^{n}$ will give very small values
of $\alpha$ in quiescence where H/R is found to decrease inward.
This will yield a very long viscous time in the inner disk
and promote outbursts that begin in the outer disk
and propagate inward. This yields model outbursts with rapid rise
and slower decline, in accord with the observations for the
optical and soft X-ray light curves of the many of the X-ray novae.
The conclusion of Vishniac and Wheeler that
the exponential luminosity decay of cooling disks is 
not compatible with models with
separate, constant values of $\alpha$ for the hot and cold states
thus constrains the nature of the outburst and suggests that for
the black hole cases, in particular, ignition of the heating on
the outside (logarithmically) is to be expected.     

More specifically, these analyses support
the conclusion that the exponential decay of the luminosity of
black hole disk systems following outbursts is consistent with a local law
for the dimensionless disk viscosity $\alpha\propto (H/R)^n$ if,
and only if, n is approximately $3/2$.
This scaling for $\alpha$ implies that disk systems
in general should exhibit approximately exponential luminosity
decay from peak luminosity whenever the hot state opacity follows a
simple power law.  
This result is apparently compatible with the internal wave
driven dynamo model for disk viscosity.  

Cannizzo \etal\ (1995) have shown that the slope of the exponential is
a function of M/$\alpha_0$, where M is the mass of the central 
object and $\alpha_0$ is the coefficient in the local viscosity
law. Cannizzo \etal\ argue that $\alpha_0\sim50$ is a universal constant.
Understanding why this coefficient differs so significantly from unity
will be a challenge for the dynamo models.  If Cannizzo \etal\ are
correct, then the slope of the exponential gives a measure 
of the mass of the central objects.  For the black hole
sources, the result is compatible with M $\sim$ 5\m and the 
conclusion that A0620-00, GS2000+25, and Nova Muscae all
have very similar black hole masses (\S 9).

These new insights into the nature of the viscosity of the
black hole sources also give a new perspective in which to ask
questions of other astrophysical sources.  The so-called
``TOADS," tremendous outburst amplitude dwarf novae 
(Howell, Szkody \& Cannizzo 1995) give evidence for especially low 
viscosity.  Another internal wave-driven mechanism, the incoherent dynamo, 
(Vishniac \& Brandenberg 1995) gives a minimum value for $\alpha_F$.
This may provide a floor opacity to the TOADS and they in turn
may provide some direct constraints on the non-local nature
of the wave-driven mechanisms.

Another problem of interest is that of disks in QSO's or AGN's.  In that 
context there is no obvious companion to generate internal waves
by tidal instability.  Some other mechanism to generate internal
waves, perhaps stellar collisions with the disk, must be sought. 

In any case, the understanding of the import of the exponential
decline of the black hole sources marks the end of a chapter
in the study of accretion disks and opens up a new range of
topics to be explored with some confidence that we do have
the first rudiments of understanding of the viscosity of
accretion disks.

\section{The Reflare}

	The ``reflare" phenomenon seems to be a common feature of
the black hole transients that show the classic fast rise and
exponential decline.  This feature calls for a physical 
explanation in general and raises the expectation that it might prove
to be an interesting physical diagnostic. 
     There have been four studies of the phenomenon of the ``reflare"
observed in the decay of many black hole X-ray novae:
Chen, Livio \& Gehrels (1993), Augusteijn, Kuulkers \& Shaham (1993), 
Mineshige (1994) and Kim, Wheeler \& Mineshige (1996b).
     
The first two studies assume that X-ray irradiation can
modulate the mass transfer rate from the companion by irradiating the 
companion during the rise, maximum and the early stage of the outburst decay
to produce the ``reflare" (mass transfer burst model).
	Chen {\it et al.} (1993) do not discuss in any detail the observable
implications of their model such as optical or X-ray light curves.  It
is not clear, for instance, what the optical response of
the disk would be, nor the degree to which even
a sharp burst of added mass will
be spread by the finite viscous response of the disk
(especially when the outer parts of the disk are in the cold state) so that
any later effect in the X-rays will be delayed with respect to
the optical and very spread out in time.  There are also questions
of whether the disk blocking they invoke to account for the
delay of the ``reflare" is consistent with their estimates
of mass transfer and energetics that depend on irradiating the
companion.  Similar issues arise in their model for the ``second maximum." 
    
	Augusteijn {\it et al.} (1993) suggest an oscillation of the light curve
in the decay in which each successive burst is a ``reflection" of 
the previous burst that heats the companion and drives more mass
transfer after some time delay.
     This model seems to be remarkably reminiscent of the recently discovered 
``mini-outbursts" in GRO J0422+32 (Callanan {\it et al.} 1995) 
and Nova Vela 1993(Bailyn {\&} Orosz 1995; 
Della Valle, Benetti, and Wheeler 1996). 
	Augusteijn {\it et al.} (1993) even
predicted bursts in Nova Per in August 1993 and December 1993,
as observed.  Augusteijn {\it et al.} (1993) deserve credit for
drawing attention to the fact that there may be some ``clock"
underlying the bursts in Nova Per and perhaps other objects, but
there are still open questions concerning their particular model.
One point that requires clarification is that Augusteijn {\it et al.} (1993)
did not clearly differentiate the ``reflare" from the
``second maximum" as we are defining them here.  They first
adjust parameters of their model to fit the ``reflare" of GS 2000+25, but
then calibrate the model on the ``second maximum" of GRO J0422+32 in
order to ``predict" the later outbursts.  It is not at all clear
that the ``reflare" and the ``second maximum" involve similar
physics since the ``reflare" is a feature only of the soft
X-rays and optical/UV and the ``second maximum" is dominated by
the power law hard X-rays, although it also has an optical component
(\S 2).
The later mini-outbursts may be related to the ``second maximum,"
as Augusteijn {\it et al.} (1993) argued, but their spectral response
is not yet well established.  They may be purely optical 
phenomena.  In addition, Augusteijn \etal\ (1993) predict
bursts of decreasing amplitude whereas,
as Bailyn \& Orosz (1995) point 
out, the observed optical bursts have nearly constant amplitude.  Augusteijn
{\it et al.} (1993) also predict a burst on 21 April, 1993 for
GRO J0422+32, and no such burst
was observed.  The models of Augusteijn {\it et al.}  (1993) also do not
consider the state of the disk, especially in its
cool, quiescent, low-viscosity state, in a self-consistent way.
     
     Unlike the pictures proposed by Chen {\it et al.} (1993) and 
Augusteijn {\it et al.} (1993), the models of Kim \etal\ (1996b)
for the optical ``reflare"
require that the outer disk and the ${\rm L_1}$ point are blocked
by the inner disk from receiving any direct irradiation
throughout the decay phase prior to the ``reflare." 
     This contradicts the  hypothesis of the X-ray-irradiated 
mass transfer burst models, that the ${\rm L_1}$ point be irradiated.
     Kim \etal\ (1996b) find the companion to intercept an angle
of about $10^{\circ}$ from the center of the disk.
    Near maximum light, the disk blocks all angles less than about $5^{\circ}$.
Kim \etal\ estimate that the companion receives
$\lta$ $10^9$ erg cm$^{-2}$ s$^{-1}$ of intercepted irradiation at maximum.
     Recent studies of the evolution of irradiated low mass companions
show that the mass transfer rate will increase as the 
companion's atmosphere expands upon heating if the irradiation fluxes 
are $\gta$ $10^{10}$ erg ${\rm {cm^{-2}}}$ $\rm {s^{-1}}$
(Podsiadlowski 1991) or
    $\gta$ $10^9$    erg ${\rm {cm^{-2}}}$ $\rm {s^{-1}}$
(Hameury {\it et al.} 1993).
    Thus the disks always shadow the L1 point and the intercepted flux
may be so low as to be unable to engender any appreciable 
structural changes in any case.

Mineshige (1994) suggests that the ``reflare" 
is caused by a sudden increase of viscosity as the disk is irradiated
and by a concurrent mass transfer burst instability.
     The origin of this sudden increase in viscosity
by a factor of 2 to 4 in extended portions of the disk, is, however, not clear.
     In the irradiated models presented of Kim \etal\ (1996b), 
there is a sudden increase of the viscosity, but it is confined to
the zone that undergoes the ``reflare."
     The viscosity parameter $\alpha$
changes by a factor $\sim$ 4.8 from $\sim$ 10 
days prior to the ``reflare" to the maximum of the ``reflare."  

The irradiated disk instability models 
of Kim \etal\ can reproduce
the optical ``reflare" observed in A0620-00. In these models
the optical ``reflare" results from an intrinsic property of the disk 
instability, the ``stagnation" phenomenon whereby the disk does not directly 
jump from the hot down to the cool state, 
but stays at an intermediate temperature due 
to an increase in the specific heat in the partially ionized matter.
Portions of the disk can then jump from this ``stagnation" state
back to the hot state giving the optical reflare.
     In the irradiated accretion disk models, the stagnation is reinforced by
the continuous heating from the irradiation reflected by the disk corona or 
chromosphere.
  This modulation of the outer disk gives no direct way to modulate
the mass flow rate in the inner disk in order to
produce a correlated soft X-ray ``reflare" as commonly observed
in these black hole candidates, a problem that plagues
other models of the ``reflare."  Kim \etal\ have shown, at least, that
the possibility of optical flaring in the outer disk
should be considered as part of the relevant physical
processes.
                     
We do not yet have a complete understanding of the physical
mechanism of the ``reflare" (or subsequent flares).
No model yet proposed can naturally account for why the
``reflare" seems to coincide with the sudden
drop in the hard X-ray flux in Nova Muscae (Miyamoto {\it et al.} 1993).
Nevertheless, these irradiated models show that effects in
the disk alone can give optical outbursts that may be
related to the optical flares seen.  They also give us
a new perspective from which to consider questions of
the irradiation of the companion.  

A critical question remains as to the frequency of occurrence
and the physical nature of the repeated flares of
systems like GRO J0422+32 and Nova Vela.  Since coverage
of these systems at late times is spotty, these 
sorts of subsequent outbursts could be common.  
The spectral coverage is even more dismal.  We have
very little idea whether these bursts are primarily
an optical phenomenon, or whether they have an
associated soft or hard X-ray component as well.
Such observational data is critical if we are to
understand whether these repeated bursts are a 
product of the outer disk, the inner disk, 
and related to the disk instability, to a tidal
instability, to a mass transfer instability,
to a combination of these processes, or to 
some other process entirely.

\section{Matching with Advective Solutions}
	
The models of Kim \etal\ (1996a,b)
give an inner region that remains in the 
hot ionized state even though the bulk of the disk drops into
a cold, low-viscosity, quiescence.  
	The cooling wave slows as the density is depleted 
in the inner disk and it does
not reach the inner edge before the next eruption occurs.
	The mass accretion rates given by these models,
$\sim10^{11}$ g s$^{-1}$ for non-irradiated models
and $\sim10^{9}$ g s$^{-1}$ for irradiated models, agrees
rather well with that deduced for A0620-00 in quiescence
by McClintock {\it et al} (1995), but the models give too low a
temperature which
McClintock {\it et al} (1995) estimate to be $\sim2\times10^6$ K.  
This problem has been discussed by Lasota (1995).
The behavior of this slow inner cooling wave may be sensitive to
the viscosity parameter and other model parameters
in a manner that needs to be explored.

A possible solution to the problem of the X-ray emission
is provided by hot flows in which quasi-radial advection
of thermal energy is a key component as reviewed by
Abramovicz in these proceedings (Narayan and Yi 1994; 
Abramovicz \etal\ 1995; Artemova \etal\ 1996).  	
Narayan, McClintock, and Yi (1996) obtain a fit to both the optical
and X-ray spectra of A0620-00 in quiescence by invoking
a hot two-temperature advective disk solution in the inner disk matched
to a steady state Keplerian disk in the outer portions that provides the 
optical luminosity.  The advective solution, however, is of
low efficiency and requires a mass flow rate of $4\times10^{14}$g s$^{-1}$,
much higher than the estimates based on steady state, geometrically
thin, optically thick disks by Marsh {\it et al.} (1994) and McClintock
{\it et al.} (1995) and much higher than the quiescent flow rates
obtained by Kim \etal\ (1996a,b).  The steady-state disks appended to
the advection solutions by Narayan, McClintock \& Yi
are not consistent with the quiescent
state of the disk being modeled.  The disk is almost surely 
not in steady state in quiescence.  
Note that, even if advection is a factor at some phases of 
quiescence,  the disk instability associated with ionization of 
hydrogen and helium and its role during outburst evolution 
can not be avoided for the conditions that are thought to exist
in quiescence.

	Lasota, Narayan, and Yi (1996) address these issues by presenting
models with a radially constant mass flow rate specified as a parameter. 
They explore conditions in which the outer cool disk can 
be in a globally stable steady state at low viscosity.  They point out that 
such configurations can be thermally unstable depending on 
the inner boundary where the advection solution takes over.
Alternatively, they might be triggered into disk instability by some
enhancement of mass transfer from the companion star.  They 
note that the disk is too cool, $\sim$2000 K, to provide the
the observed quiescent optical luminosity, an aspect common to all
disk instability models for these systems (Huang \& Wheeler 1989;
Mineshige \& Wheeler 1989; Mineshige, Kim \& Wheeler 1990).  

	One obvious possibility is that the blue optical flux observed in
quiescence of A0620-00 is that of the hot spot generated
by the transfer stream that is, to first order,
constant during the whole outburst process. 
     A0620-00, GS 2000+25, and Nova Muscae are strikingly similar objects
in terms of their light curves.
     They also have similar orbital period and mass function, which suggests
they have similar accretion disks.
     Eclipse mapping of GS 2000+25 in quiescence shows 
evidence for a hot spot which is much brighter than the disk itself
(Casares, Charles \& Marsh 1995).
     The bright spot in GS 2000+25 
clearly implies that there is continuous mass transfer in quiescence, 
the major assumption of the disk instability model.  The fact that the
bright spot dominates the disk luminosity is consistent
with the cold disk we predict in quiescence.
	The possibility that the hot spot contributed substantially
to the quiescent optical emission of A0620-00  was discussed by
McClintock {\it et al.} 1983).  The optical/UV observations reported
by McClintock, Horne, and Remillard (1995) show that the bolometric flux
detected by HST is consistent with that expected from the 
transfer stream.  They fit the continuum spectrum of the ``disk" with
a reddened 9000 K black body and for a distance of 1 kpc and derive 
an emitting area of $1.2\times10^{20}$ cm$^2$, 80 times smaller than
the disk.  The implied bolometric luminosity of this emission, assuming
an optically thick sphere with this area and effective temperature,
is $4.6\times10^{31}$ erg s$^{-1}$.  The rate of release of
potential energy by the transfer stream striking the outer edge of 
the disk at radius R$_d$ 
is given approximately by $1.3\times10^{30}M{\rm {\dot M_{15}}}/R_{d,11}$
erg s$^{-1}$ where M is the mass of the compact object in solar masses,
${\rm {\dot M_{15}}}$ is the secondary transfer rate in $10^{15}$ g s$^{-1}$
and  R$_{d,11}$ is the outer disk radius in units of $10^{11}$ cm.  Clearly,
for M${\rm {\dot M_{15}}}\sim30$, a plausible case could be made
that all the optical luminosity arises in a hot spot and not in the
disk proper.  It is not clear whether ascribing all the optical
luminosity to the hot spot is consistent with the lack of observed 
orbital modulation (McClintock, Horne \& Remillard 1995). 

There is still the question of the quiescent X-ray emission from
A0620-00 which can be matched by the advection solutions with
appropriate choices of parameters, but not with the present 
disk instability models.
The advection solutions require some means of
severely depleting the surface density of the inner portions
of the disk as the disk approaches quiescence.  It is difficult
to see how such a solution matches physically in terms of
the surface density and angular momentum with the outer
thin quiescent disk.  Perspective on this problem can be
gained by considering some time scales inherent in the advection
solution. The high temperature, low efficiency advection solutions
require a very low density and hence a low mass in the advection region.
The published solutions assume an {\it ad hoc} supply of mass 
from the Keplerian disk to the advection region in order to
maintain this steady-state condition (Narayan, McClintock and Yi 1996; Lasota, 
Narayan, and Yi 1996).  The disk instability models, on the other hand,
give a much lower mass flow rate in quiescence.  It is thus of interest
to inquire how long the advective solution would survive if the mass
flow were cut off.  Using the expression for the density distribution
given in Narayan, McClintock, and Yi and the parameters for their fit
to A0620-00 ($\alpha = 0.1$, M = 4.4$M_{\odot}$, ${\rm {\dot M_{15}}}=0.4$), 
the mass is $1.9\times10^2$R$_2^{3/2}$ g.  For the derived
mass flow rate in the advective region, the time to consume the
advective region in the absence of mass input is 
M/${\rm {\dot M}}$=14 s $R_{a,9}^{3/2}$, where $R_{a,9}$ 
is the boundary between the advective region and Keplerian disk
in units of $10^9$ cm.  Clearly, the advection region 
is very fragile and must be constantly 
supplied at the high mass flow rates required by the low radiative
efficiency in order to be sustained.  This is not a trivial requirement. 

The high mass flow rate necessary to maintain the advective region, 
${\rm {\dot M_{15}}}=0.4$, does not
arise naturally in the quiescent disk instability models.  It 
must, rather, be supplied from the matter in the Keplerian disk
by some sort of ill-understood ablation process.  One can at least
ask what the implications are of supplying mass at this rate.
To determine this, we fit an approximate power law to the quiescent
disk structure of the models of Kim \etal\ (1996a,b), 
recognizing that this will
be different for different viscosity prescriptions.  For the
quiescent mass distribution corresponding to Figure 5, we find   
$$
	{dm\over dr}~=~{1\over v_a}{dm\over dt}\sim10^{13}~r_{10}^{1.4}~g,
					\eqno (5)
$$
where v$_a$=dR$_a$/dt is the speed with 
which the boundary of the advection region
advances into the Keplerian disk and dm/dt is the rate of ablation
necessary to feed the advection region.  
This implies a characteristic time scale to consume the disk of 
$$
	\tau~=~{R_a\over v_a}~=~2.5\times10^8~R_{a,10}^{2.4}~s,
					\eqno (6)
$$
where R$_{a,10}$ is the outer boundary of the advection region 
in units of 10$^{10}$ cm.
	The advection solution of Narayan, McClintock \& Yi (1996)
that reproduces the ROSAT X-ray observations
cannot remain with an outer boundary of order 10$^9$ cm for times comparable
to the recurrence time.  If such an advection region is fed by
ablation from the Keplerian disk 20 years after the outburst, the
boundary must be in excess of 10$^{10}$ cm.  This seems to be a rather
extreme criterion.
	An alternative is that the quiescent X-rays do not arise
in the inner disk at all.  The observed flux is low and it is
possible that it arises in flare activity on the companion.   
 
Clearly more work is needed to understand the state of the
disk in both quiescence and outburst.  The black hole
X-ray transients are excellent laboratories for these
studies and the disk instability models give a framework
in which to pose relevant questions.

\section{The Hard Power Law, Radio Outbursts, and Positrons}

The hard power law component that is frequently observed
in black hole X-ray transients is very similar to that
observed in AGN as discussed in these proceedings by Svensson.
Since at some level we know more about the transient systems --
the definite existence of accretion disks, orbital parameters
and, in the best cases, masses of the components -- we
may, with some hubris, expect that these transient systems
may help us to understand the hard radiation mechanisms of AGN,
rather than the other way around.

It is commonly assumed that the power law component represents a 
Comptonized thermal spectrum arising in a hot plasma
that is itself part of a coronal structure that is in hydrostatic
equilibrium.  Such a simple model for the emission can
fit some objects at some epochs, but that does not make
it unique or correct.  Such models ignore
the obvious evidence for non-thermal particles and
magnetic fields implied by the common radio outbursts
that are frequently associated with the X-ray bursts
(Han and Hjellming 1992).
The recent super-luminal sources are only the most
extreme example.  It is most likely that the non-thermal
particles and magnetic fields arise in the disk and
hence must be incorporated into models of the hard 
power-law emission.  
Kusunose has shown that thermal models are not unique.
Non thermal models can produce power law emission
that has a constant slope independent of luminosity,
as observed (Kusunose \& Mineshige 1995).
In addition, there is ample 
evidence for outflow at least at some epochs and so
there is reason to question the assumption of
hydrostatic equilibrium for the plasma. 

The soft X-ray component that probably arises in the accretion
disk peaked more slowly than the hard power law flux in
Nova Muscae.  This may mean that the inner disk was incomplete
in quiescence or the early phase of the outburst.  The radius
of the geometrically thin, optically thick disk may have
shrunk in response to increased mass flow attendant with
the disk instability in the outer disk, thus giving rise
to the delayed rise of the soft flux.
It may be that in systems like V 404 Cyg and GRO J0422+32,
the inner disk never properly forms to give soft X-rays,
for reasons that are not currently understood.

The first flare of the hard flux in systems like Nova Muscae
may be better associated, not with a static, thermal plasma,
but with a non-thermal, magnetic, pair-rich
outflow (Moscoso \& Wheeler 1994).  This phase shows
QPO's, correlated radio synchrotron bursts, and at least
in Nova Muscae, the line feature that is plausibly associated
with annihilation.  If this is the annihilation line,
then it is much too narrow to represent annihilation
in the region where positrons are created and hence
implies flow of some kind from the location of the creation
of the positrons to the locale of their annihilation.

The ``second maximum," as defined here is primarily
characterized by a hard power law spectrum and the virtual 
absence of a soft X-ray component from the inner disk.
This feature may arise in a structure that more
closely resembles a quasi-static corona of the sort 
frequently modeled in the literature (Liang 1990).  The fact that
the disk component of the soft X-ray flux declines
as this late hard component comes up strongly suggests
that the corona is displacing the geometrically
thin disk.  With a larger effective inner radius, the
disk simply becomes too cool to emit soft X-rays.

There is some speculation that the ``second maximum" may be related
to an advection solution, but the problems associated
with the time scales and the feeding of such a region by
the outer thin disk outlined in the previous section may again apply.   
There is also some question of whether this feature
could be related to a tidal instability that is triggered
when the disk radius grows beyond the 3 to 1 resonance
due to the onset of the disk instability.  The long time
delay, 200 days, might be accounted for by the 
slow linear growth time of the instability (Ichikawa, Mineshige,
\& Kato 1994).  It is
not clear that this picture can naturally account for
the X-ray spectral characteristics of the ``second maximum." 

To better understand the possible orgin of the power law
in dynamical environments and the nature of the ``annhilation
feature" in Nova Muscae, Moscoso, Kusunose, and Wheeler
are constructing a model to explore the
source of the outflow in the primary outburst.  This
model consists of an inner hot, pair-rich corona represented
by a single zone.  Above this corona, photon annihilation
will generate electron/positron pairs and associated 
annihilation.  The parallel component of the average momentum of
the photons that produce pairs is assumed to represent the
bulk outflow momentum of pairs.  The remaining momentum is
randomized to provide the 
thermal component of the pair energy.  Account will be
taken of both the isotropic and anisotropic Comptonization.
This simple model will give an estimate of the typical
flow time scales, speeds, and the optical depth so
that annihilation line profiles can be estimated.  

\section{Evolution}

Many fascinating questions arise with respect to the evolution
of these black hole transients.  These involve the
large mass ratios, the suggestion that the black holes
may be rather moderate in mass, and the significance
of lithium in some of them.

With typical primary masses of 5\m and secondary masses of
0.5\m, the mass ratio of these systems, $\sim 10$, is
very large. Most massive binaries are thought to have 
mass ratios closer to unity, but there are clearly strong
selection effects that bias observations away from such
large ratios.  Nevertheless if this is merely an accident of birth, 
it seems odd that it is the future black hole systems that
favor small mass companions.

It is not clear how much the secondary may have altered during
the evolution of the system (De Kool, van den Heuvel \& Plyser 1987).  
Many of the secondaries seem
to be at least slightly evolved, the secondary in V404 Cyg
definitely so (Casares, Charles \& Naylor 1992). 
The atmospheres, where they can be observed
in quiescence, are, with some small exceptions, characteristic
of main sequence stars, so they can not be highly evolved.
They may have lost mass not only to Roche lobe overflow,
but also to the process of being blasted by X-ray radiation
at recurrent intervals since the black hole formed
(Mineshige, Kim \& Wheeler 1990).
The small orbital periods that characterize these systems,
suggest that they have been subject to common envelope evolution
and ejection, but the low mass companions have little leverage
on the envelope of the black hole progenitor and it is
not clear how such systems could escape core coalescence.
There is some speculation that the secondary was not
primordial, but formed out of the envelope of the massive
star by some adjunct process to creating the black hole
(Podsiadlowski 1996).

There is increasing evidence that there are upper limits
as well as lower limits to the masses of the black holes
in these systems (see, eg, Chen, Shrader \& Livio (1996) for
a compilation).  The upper mass limits are interestingly low,
perhaps systematically substantially less than 10\m in many cases.
Rather tight limits have been derived for Nova Oph 1977
(4.9$\pm$1.3\m; Remillard \etal\ 1996), GS 2000+25
(5.9 - 7.5 \m; Filippenko, Matheson and Barth 1995), GRO J0422+32
(3.57$\pm0.34$; Filippenko, Matheson, and Ho 1995),
and Nova Sco (4.0 - 5.2 \m; Bailyn \etal\ 1995).  Nova
Muscae has been estimated to be in the range 4 - 8 \m (Chen \etal\ 1996).
If a temporary eclipse by the disk was seen by Haswell \etal\ (1993)
an upper limit of 5 \m can be assigned to A0620-00.
V 404 Cyg is the sole current exception with a mass that
might plausibly be as large as 12 \m (Shahbaz \etal\ 1994;
but see Sanwal \etal 1996 for an upper limit).

The possible clustering of masses around 5 \m would
be entirely consistent with the notion that the
decline rate is a measure of the mass of the compact
object (divided by $\alpha_0$; \S 5).  In this interpretation,
the commonly observed decline rate with an e-fold time
of about 30 days would signify a common mass.  An
important caveat to this conclusion is the evidence that
some neutron star systems also display a decline rate
of nearly 30 days (Chen \etal\ 1996). 

These masses suggest that the black holes did not simply
form by the collapse of a massive star.  Clearly the collapse
of an entire massive star would give far larger masses.
A more reasonable possibility is that the collapse only
involved the core of the star, the envelope having been
ejected in a wind or common envelope process. As a rule of 
thumb, the core of a massive star is about 1/3 the original
main sequence mass.  Since moderately massive stars 
are commonly presumed to give rise to supernovae
and neutron stars, the progenitors of black holes might
be expected to have main sequence masses in excess of
30 \m (Shields and Wheeler 1973).  
This would make a core mass in excess of 10 \m, 
still too massive to account for the low estimates 
given above.   There may be a need for some other 
evolutionary process.

One possibility is that the object that is now a black
hole began its compact life as a neutron star.  If there
was a stage of rapid mass transfer, then the neutron
star might have been swamped so that an envelope formed
around it.  The resulting object with a neutron star in
the center and a quasi-static envelope supported by
energy from the gravitational energy or nuclear burning of accreted
matter is known as a Thorne-\.Zytkow Object (Thorne
and \.Zytkow 1977).  In such a case the neutron star might
continue to accrete until it surpassed the stable mass limit
and then collapse to form a black hole and consume the
remainder of the envelope.  The mass of the resulting
black hole would be the sum of the neutron star plus
the envelope and could be rather modest.  This scenario
would demand that the secondary lose substantial mass.
The secondary would have to be evolved, as seems to
be frequently the case,
although it could retain a thin H blanket.
This possibility would have to be reconciled with the atmospheric 
observations.  Forming a Thorne-\.Zytkow Object rather than burying both
stars in a common envelope would also presumably require
special conditions.  The possibility that the original 
star is consumed, but the current secondary spun off
to coalesce anew is a complication that might be worth
considering.

One interesting connection of Thorne-\.Zytkow Objects with
the current population of  black holes is lithium.
The two black hole candidates that have sufficiently bright
quiescent luminosity to be examined, V 404 Cyg (Martin \etal\ 1992) 
and AO620-00 (Marsh \etal\ 1994), 
have both revealed excess of lithium in 
the atmospheres of the companions.  This observation
cannot be a unique signature of black holes systems
because lithium has also been detected in the neutron
star system Cen X-4 (Martin \etal\ 1994), but it is nevertheless intriguing.
No such lithium enrichment has been observed in 
nova-like systems containing white dwarfs (Martin \etal\ 1995).   

The lithium is short-lived in convective stellar
atmospheres, so the strong presumption is that it
must be freshly created in these systems, not primordial.
The most likely origin is by spallation in the atmosphere
of the secondary by fast particles accelerated by processes near the
black hole (or neutron star).  Another possibility
is that the lithium is created in the disk, where
spallation is, if anything, more likely, but then
the problem exists of preserving and transporting
the lithium to the secondary star where it is observed.
Yet another possibility, more remote, but perhaps not
impossible, is that the lithium
is associated with a Thorne-\.Zytkow Object precursor
phase.  Under some circumstances, the deep convective
burning in a Thorne-\.Zytkow Object can produce lithium
(Cannon 1993; Biehle 1994; Podsiadlowski, Cannon, \& Rees 1995).
It is not clear that the modest mass Thorne-\.Zytkow Objects
that would be the precursors of the modest mass
black holes would have this property (this is perhaps more
likely with the large mass black hole of V 404 Cyg),
nor how the lithium could be transferred to and
preserved in the atmosphere of the secondary.   

The observation of the lithium excess also couples to
one more conundrum in the black hole systems.  The
putative annihilation line in Nova Muscae and
the corresponding broad feature that has appeared in
1E1740.7 (the ``Einstein Source") are, within
the errors, centered at about 480 Kev, not 511 Kev.
This has been interpreted as requiring that the
annihilation occur sufficiently deeply in a gravity
well that the shift can be attributed to a gravitational
redshift (Chen, Livio \& Gehrels 1993; Moscoso \& Wheeler 1994).  
As it happens, one of the principle $\gamma$-ray de-excitation
lines of lithium formed in spallation is at 483 Kev.
This coincidence has given rise to considerable speculation
that the ``annihilation feature" is actually de-excitation
of the newly forming lithium observed in the companion
(Martin \etal\ 1992, 1994; Martin, Spruit, and van Paradijs 1994).  
Although there are some numerical 
coincidences, no quantitative model of spallation of
the lithium has been provided to account for the
observations.  One counter
argument is that there should be other associated 
spallation $\gamma$-rays that are not observed and a
counter-counter argument is that there may be some
hint, in Nova Muscae, that the 480 Kev feature has
a double peak.  It will take much better signal to 
noise to resolve that issue.  The 480 Kev feature in
Nova Muscae was accompanied by another, weaker line
at 175 Kev which can be kinematically accounted for
by a blue-shifted back scattering of the annihilation
line (Hua and Lingenfelter 1993).  It is not clear that a lithium model can
account for this ``back-scatter" feature. Other 
objections to the lithium interpretation are that
the 480 Kev feature, while too narrow to arise
in the region where positrons are created, has
a finite width in both Nova Muscae and the
1E source.  This means that the feature is unlikely
to arise in the atmosphere of the secondary
even though that is where the optical line of
lithium is seen.  The accretion disk is a more
likely locale to produce such broad de-excitation 
lines by thermal or Doppler broadening.  One might
then have the inverse problem of the annihilation line 
interpretation, because if the line is from 
lithium, then it can suffer very little or
no gravitational redshift.   Clearly the interpretation
of this feature as either lithium or positronium
requires more careful quantitative analysis. 
Yet another possibility is that circumstances contrive
to give the ``line" as a result of continuum scattering
in an appropriately beamed flow (Skibo, Dermer \& Ramaty 1994).

\section{Conclusions}

The basic picture in which a companion star feeds mass 
at nearly a constant rate into an accretion disk
that is susceptible to an ionization limit-cycle
instability is still the most plausible explanation
for the black hole X-ray transients.  The characteristic 
light curve with a fast rise and exponential decline is
consistent (even demands) that the instability begins
in the outer portion, logarithmically, of the disk.
In such a case, the optical emission should rise before
any hard emission.  This is difficult to check with
current search techniques, but something like an
all sky optical monitor could provide critical constraints
on this important qualitative prediction.

As the heating wave propagates inward and leads to
a dramatic increase in the flow rate in the inner
disk, one might expect outflows to be generated.
These outflows could then, in turn, be related to
the ubiquitous radio bursts including, in the most
extreme cases, the superluminal sources, and
the occasional observation of annihilation lines.
If the outflow region interdicts the inner portion
of the accretion disk so that the optically thick,
geometrically thin disk can not, at first,
extend down to the last stable circular orbit,
then the disk may be too cool to emit X-rays.
This may be why the hard flux arises earlier
than the soft flux in Nova Muscae despite
the prediction that the instability should
begin in the outer disk.

In this picture, the hard power law flux associated
with the onset of primary outburst in the 
fast-rise, exponential decline systems, should be
associated with an outflow, probably pair rich and
magnetic, not in a quasi-static corona, nor
even in a quasi-radial advective region.  An important
task is to develop physical models that can help
to discriminate these various possibilities.
This is the stage at which systems
like Nova Muscae display the strongest QPO's
and hence one is invited to examine models for
QPO's that do not directly involve only the
disk or a radial inflow, but a dynamic outflow.

After a period of order a few days after the rise of the
hard flux, the disk may settle into a quasi-steady
declining state where it can emit copious soft X-rays and
the hard power law source, perhaps in the outflow,
declines rather more sharply.  The exponential
decline of the soft X-rays is a critical constraint
on the physical nature of the disk.  It is completely
consistent with the predictions of the disk instability
model, demands that a cooling wave be propagating in
the disk during the decline, and is consistent
only with a tightly-constrained prescription of the
disk viscosity.  There is only one self-consistent
model for disk viscosity in the literature that 
fulfils these constraints, that based on a magnetic
dynamo driven by internal waves.  

The common ``reflare" phenomenon during the exponential
decline may be related to the self-irradiation
of the disk.  The fact that it occurs in Nova Muscae
as the hard component reaches a minimum may be a clue
that requires pursuing.  It is not at all clear that
this feature is related to a modulation of the mass transfer
from the companion. 

The common ``second maximum" is dominated by a hard power law
source in Nova Muscae rather than the disk thermal radiation,
and presumably also in other similar systems.  This phase
tends not to be accompanied by QPO's.  The power law
source in this phase may arise in a quasi-static corona
of a more traditional sort that can arise as mass
flow rates decline.  Although the spectral index
is similar, this phase may thus involve a very different
physical environment than the first epoch of hard power
law emission that is dominated by outflow.  The formation
of this coronal region in the inner disk may again
interdict the inner optically thick disk so that the emission
of soft X-rays is inhibited.  This could account for the
precipitous decline in the disk component of the soft
X-rays in Nova Muscae as the hard power law source
grows to dominate the ``second maximum."  The ``corona"
associated with the ``second maximum" may be related to
an advective flow, but the current advective models
have been fit to the much later, currently observed quiescent phase
of AO620-00 and it is not clear that they can
naturally account for the transient behavior of the
``second maximum."  At any phase there are some difficult
questions to be addressed as to how the relatively high
mass flow rates of the advective solutions can be
stably fed by the naturally low mass flow rates
in the quiescent Keplerian disk.

The black hole X-ray transients provide a continuing source
of stimulating physics problems worthy of Igor Novikov.
Their time-dependence and multi-component spectra, while 
a complication, may yet
yield the sort of direct irrefutable evidence that we are dealing
with black holes that is one of the holy grails of
modern astrophysics.  As a practical matter, this
would allow us to search for and identify black holes
of stellar mass in a way that is not possible when
one is restricted to argument by mass function that
``it is too massive to be a neutron star."  Finally,
with the lessons learned in these marvelous laboratories,
we will be more strongly armed to attack the
stubborn problem of the nature of QSO's and AGN's
and their supermassive black holes.

\begin{acknowledgments}      
I would like to thank all my colleagues who have taught me
so much about black hole systems and accretion disks,
especially John Cannizzo, Pranab Ghosh, Min Huang, Shin Mineshige, 
Soon-Wook Kim, Michael Moscoso, Greg Shields, Rob Robinson,
Ethan Vishniac, and Masaaki Kusunose, 
with whom I have had the pleasure of working
at Texas on these issues.
This work has supported in part by NASA 
through grants NAGW-2975 and NAG5-3079.
\end{acknowledgments}



\begin{thebibliography}{}

\bibitem[]{}
{\sc Abramowicz, M. A., Chen, X., Kato, S., Lasota, J.-P., \& Regev, O.} 1995,
	{\it ApJ}, {\bf 438}, L37   

\bibitem[]{}
{\sc Artemova, I. V., Bisnovatyi-Kogan, G. S., Bj\"ornsson, G.,
	\& Novikov, I. D.} 1996, {\it ApJ}, in press 

\bibitem[]{}
{\sc Augusteijn, T., Kuulkers, E., \& Shaham, J.} 1993, {\it A\&A}, 
	{\bf 279}, L13 

\bibitem[]{}
{\sc Bailyn, C. D. {\it et al.}} 1995, {\it Nature}, {\bf 374}, 701

\bibitem[]{}
{\sc Bailyn, C. D. {\&} Orosz, J. A.} 1995, {\it ApJ}, {\bf 440}, L73

\bibitem[]{}
{\sc Biehle, G. T.} 1994, {\it ApJ}, {\bf 420}, 364    

\bibitem[]{}
{\sc Blandford, R. D., and Znajek, R. L.} 1977, {\it MNRAS}, {\bf 179}, 433    

\bibitem[]{}
{\sc Callanan, P. J. {\it et al.}} 1995, {\it ApJ}, {\bf 441}, 779    

\bibitem[Cannizzo\ 1993a]{C93a} 
{\sc Cannizzo, J.K.} 1993a, in {\it Accretion Disks in Compact Stellar Systems},
	ed. J.C. Wheeler (World Scientific Press: Singapore), 6

\bibitem[Cannizzo\ 1993b]{C93b}
{\sc Cannizzo, J.K.} 1993b, {\it ApJ}, {\bf 419}, 318

\bibitem[Cannizzo\ 1994]{C94}
{\sc Cannizzo, J.K.} 1994, {\it ApJ}, {\bf 435}, 389

\bibitem[Cannizzo, Chen, \& Livio\ 1995]{CCL95}
{\sc Cannizzo, J.K., Chen, W., \& Livio, M.} 1995, {\it ApJ}, 
	{\bf 454}, 880 

\bibitem[]{}
{\sc Cannizzo, J. K., Ghosh, P., \& Wheeler, J. C.} 1982, {\it ApJ}, 
	{\bf 260}, L83

\bibitem[]{}
{\sc Cannon, R. C.} 1993, {\it MNRAS}, {\bf 263}, 817

\bibitem[]{}
{\sc Casares, J. C. \& Charles, P. A.} 1994, {\it MNRAS}, {\bf 271}, L5

\bibitem[]{}
{\sc Casares, J. C., Charles, P. A., \& Naylor, P. A.} 1992, 
	{\it Nature}, {\bf 355}, 614

\bibitem[]{}
{\sc Chen, W., Livio, M., \& Gehrels, N.} 1993, {\it ApJ}, {\bf 408}, L5

\bibitem[]{}
{\sc Chen, W., Shrader, C. R., \& Livio, M.} 1996, {\it ApJ}, in press

\bibitem[]{}
{\sc Cheng, F. H., Horne, K., Panagia, N., Shrader, C. R., Gilmozzi, R.,
	Paresce, F., \& Lund, N.} 1992, {\it ApJ}, {\bf 396}, 664

\bibitem[]{}
{\sc Crary, D. J., Kouveliotou, C., van Paradijs, J., van der Hooft, F.,
	Scott, D. M., Paciesas, W. S., van der Klis, M., Finger, M. H.,
	Harmon, B. A., \& Lewin, W. H. G.} 1996, {\it ApJ}, in press

\bibitem[]{}
{\sc de Kool, M., van den Heuvel, E. P. J., and Pylyser, E.} 1987,
	{\it A\&A}, {\bf 183}, 47 

\bibitem[]{}
{\sc Della Valle, M., Benetti, S. \& Wheeler J. C.} 1996, 
	{\it A\&A}, in press   

\bibitem[]{}
{\sc Ebisawa, K. \etal} 1994, {\it PASJ}, {\bf 46}, 375                                        
\bibitem[]{}
{\sc Ebisawa, K.} 1996, private communication

\bibitem[]{}
{\sc Faulkner, J.,  Lin, D. N. C., \& Papaliozou, J.} 1983, {\it MNRAS},
	{\bf 205}, 359

\bibitem[]{}
{\sc Filippenko, A. V.,  Matheson, T., \& Ho, L. C.} 1995, {\it ApJ}, 
	{\bf 455}, 614  

\bibitem[]{}
{\sc Filippenko, A. V.,  Matheson, T., \& Barth, A. J.} 1995, {\it ApJ},
        {\bf 455}, L139 

\bibitem[]{}
{\sc Fishman, G. J. \etal} 1989, in
{\it Gamma Ray Observatory Science Workshop}, Eds. C. R. Shrader, 
	N. Gehrels, \& B. Dennis, (CP-3137, NASA: Greenbelt, MD)  p239
                                                                
\bibitem[]{}
{\sc Ford, H. C., Harms, R. J., Tsvetanov, Z. I., Hartig, G. F., 
	Dressel, L. L., Kriss, G. A., Bohlin, R. C., Davidsen, A. F.,
	Margon, B, \& Kochhar, A. K.} 1994, {\it ApJ}, {\bf 435}, L27

\bibitem[]{}
{\sc Fukue, J.} 1992, {\it PASJ}, {\bf 44}, 663

\bibitem[]{}
{\sc Goldwurm, A. {\it et al.}} 1992, {\it ApJ}, {\bf 389}, L79

\bibitem[]{}
{\sc Han, X., \& Hjellming, R. M.} 1992, {\it ApJ}, {\bf 400}, 304

\bibitem[]{}
{\sc Harmon, B. A. {\it et al.}} 1994, {\it Second Compton Symposium}, 
	Eds. C. E. Fichtel, N. Gehrels \& J. P. Norris (AIP: New York), 210

\bibitem[]{}
{\sc Harmon, B. A. {\it et al.}} 1995, {\it Nature}, {\bf 374}, 703

\bibitem[]{}
{\sc Haswell, C. A. {\it et al.}} 1993, {\it ApJ}, {\bf 411}, 802

\bibitem[]{}
{\sc Howell, S. B., Szkody, P., \& Cannizzo} {\it ApJ}, {\bf 389}, 337   

\bibitem[]{}
{\sc Huang, M., \& Wheeler, J. C.} 1989, {\it ApJ}, {\bf 343}, 229

\bibitem[]{}
{\sc Hua, X.-M. \& Lingenfelter, R. E.} 1993, {\it ApJ}, {\bf 416}, L17

\bibitem[]{}
{\sc Hjellming, R. M.} 1996, private communication 

\bibitem[]{}
{\sc Hjellming, R. M. \& Rupen, M. P.} 1995, {\it Nature}, {\bf 375}, 464

\bibitem[]{}
{\sc Ichikawa, S., Mineshige, S., \& Kato, T.} 1994, {\it ApJ}, {\bf 435}, 748

\bibitem[]{}
{\sc Kim, S.-W., Wheeler, J. C., \& Mineshige, S.} 1996a, {\it ApJ}, submitted

\bibitem[]{}
{\sc Kim, S.-W., Wheeler, J. C., \& Mineshige, S.} 1996b, {\it ApJ}, submitted


\bibitem[]{}
{\sc Kouveliotou, C., Finger, M. H., Fishman, G. J., Meegan, C. A.,
	Wilson, R. B., Paciesas, W. S., Minamitani, T., \& van Paradijs, J} 
	1993 in {\it Compton Gamma Ray Observatory},  
	Eds. M. Friedlander, N. Gehrels, \& D. J. Macomb (AIP: New York), 319 

\bibitem[]{}
{\sc Kusunose, M., \& Mineshige, S.} 1995, {\it ApJ}, {\bf 440}, 100  

\bibitem[]{}
{\sc Lasota, J.-P.} 1995 in {\it Compact Stars in Binaries; Proceedings
	of IAU Symposium 165},  Eds. E. P. J. van den Heuvel and 
	J. van Paradijs (Kluwer: Dordrecht), in press      

\bibitem[]{}
{\sc Lasota, J.-P., Narayan, R., \& Yi, I.} 1996, submitted to {\it A\&A}

\bibitem[]{}
{\sc Liang, E. P.} 1990, {\it A\&A}, {\bf 227}, 447

\bibitem[]{}
{\sc Ling, J. C., Mahoney, W. A., Wheaton, W. A., \& Jacobson, A. S.} 1987,
{\it ApJ}, {\bf 321}, L117  

\bibitem[]{}
{\sc Marsh, T. R., Robinson, E. L., \& Wood, J. H.} 1994, 
	{\it MNRAS}, {\bf 266}, 137

\bibitem[]{}
{\sc Martin, E. L., Casares, J.,  Charles, P. A., \& Rebolo, R.} 1995, 
	{\it A\&A}, {\bf 303}, 785

\bibitem[]{}
{\sc Martin, E. L., Rebolo, R., Casares, J. \& Charles, P. A.} 1992, 
	{\it Nature}, {\bf 358}, 129

\bibitem[]{}
{\sc Martin, E. L., Rebolo, R., Casares, J. \& Charles, P. A.} 1994, 
	{\it ApJ}, {\bf 435}, 791

\bibitem[]{}
{\sc Martin, E. L., Spruit, H. C., \& van Paradijs, J.} 1994, 
	{\it A\&A}, {\bf 291}, L43


\bibitem[]{}
{\sc McClintock, J. E., Horne, K., \& Remillard, R. A.} 1995, {\it ApJ}, 
	{\bf 442}, 358

\bibitem[]{}
{\sc McClintock, J. E., Petro, L. D., Remillard, R. A., \& Ricker, G. R.} 1983,
     {\it ApJ}, {\bf 266}, L27

\bibitem[]{}
{\sc McClintock, J. E., \& Remillard, R. A.} 1986, {\it ApJ}, {\bf 308}, 110

\bibitem[]{}
{\sc Meyer, F. \& Meyer-Hofmeister, E.} 1981, {\it A\&A}, {\bf 104}, L10

\bibitem[]{}
{\sc Mineshige, S.} 1994, {\it ApJ}, {\bf 431}, L99

\bibitem[]{}
{\sc Mineshige S., Kim S.-W., \& Wheeler, J. C.} 1990, {\it ApJ}, {\bf 358}, L5

\bibitem[]{}
{\sc Mineshige, S., \& Osaki, Y.} 1983, {\it PASJ}, {\bf 35}, 377

\bibitem[]{}
{\sc Mineshige, S., Tuchman, Y., \& Wheeler, J. C.} 1990, {\it ApJ}, 
{\bf 359}, 176

\bibitem[Mineshige \& Wheeler\ 1989]{MW89}
{\sc  Mineshige, S., \& Wheeler, J.C.} 1989, {\it ApJ}, {\bf 343}, 241

\bibitem[Mineshige, Yamasaki \& Ishizaka\ 1993]{MYI93}
{\sc  Mineshige, S., Yamasaki, T., \& Ishizaka, C.} 1993, {\it PASJ}, 
	{\bf 45}, 707

\bibitem[]{}
{\sc Mirabel, I. F., \& Rodriguez, L. F.} 1994, {\it Nature}, {\bf 371}, 46

\bibitem[]{}
{\sc Miyamoto, S. {\it et al.}} 1993, {\it ApJ}, {\bf 403}, L39 


\bibitem[]{}
{\sc Miyoshi, M., Moran, J., Herrnstein, J., Greenhill, L., 
	Nakai, N., Diamond, P., \& Inoue, M.} 1995, {\it Nature}, 
	{\bf 373}, 127

\bibitem[]{}
{\sc Moscoso, M., \& Wheeler, J. C.} 1994, in {\it Interacting Binary Stars},
     Astronomical Society of the Pacific Conference Series Vol. 56,
     ed. A. W. Shafter (San Francisco: Astronomical Society of the Pacific), 100

\bibitem[]{}
{\sc Narayan, R., \& Yi, I.} 1994,  {\it ApJ}, {\bf 428}, L13  

\bibitem[]{}
{\sc Narayan, R., McClintock, J. E., \& Yi, I.} 1996, {\it ApJ}, {\bf 457}, 821

\bibitem[]{}
{\sc Podsiadlowski, Ph.} 1991, {\it Nature}, {\bf 350}, 136

\bibitem[]{}
{\sc Podsiadlowski, Ph.} 1996, private communication

\bibitem[]{}
{\sc Podsiadlowski, Ph., Cannon, R. C., \& Rees, M. J.} 1995, {\it MNRAS}, 
	{\bf 274}, 485

\bibitem[]{}
{\sc Remillard, R. A., McClintock, J. E., \& Bailyn, C. D.} 1992, {\it ApJ}, 
	{\bf 399}, L145

\bibitem[]{}
{\sc Remillard, R. A., Orosz, J. A., McClintock, J. E., \& Bailyn, C. D.} 
	1996, {\it ApJ}, {\bf 459}, 226

\bibitem[]{}
{\sc Sanwal, D., Robinson, E. L., Zhang, E., Colom\'e, C., Harvey, P. M.,
	Ramseyer, T. F., Hellier, C., \& Wood, J. H.} 1996, {\it ApJ}, 
	{\bf 460}, 437

\bibitem[]{}
{\sc Shahbaz, T., Ringwald, F. A, Bunn, J. C., Naylor, T., 
	Charles, P. A., \& Casares, J.} 1994, {\it MNRAS}, {\bf 271}, L10

\bibitem[Shakura \& Sunyaev\ 1973]{SS73}
{\sc Shakura, N.I., \& Sunyaev, R.A.} 1973, {\it A\&A}, {\bf 24}, 337

\bibitem[]{}
{\sc Shields, G. A. \& Wheeler, J. C.} 1973, {\it Nature}, 259, 642

\bibitem[]{}
{\sc Skibo, J. G., Dermer, C. D., \& Ramaty, R.} 1994, {\it ApJ}, 431, L39

\bibitem[]{}
{\sc Smak, J.} 1982, {\it Acta Astr}, {\bf 32}, 199

\bibitem[]{}
{\sc Smak, J.} 1983, {\it ApJ}, {\bf 272}, 234


\bibitem[]{}
{\sc Smak, J.} 1984, {\it PASP}, {\bf 96}, 1

\bibitem[]{}
{\sc Sunyaev, R. A. {\it et al.}} 1992, {\it ApJ}, {\bf 389}, L75

\bibitem[]{}
{\sc Tanaka, Y.} 1990, in {\it the 23rd ESLAB Symposium 
    on Two Topics in X-ray Astronomy},
   Ed. N. E. White (ESA SP$-$296), p2

\bibitem[]{}
{\sc Tanaka, Y., \& Lewin, W. H. G.} 1995, in {\it X-Ray Binaries}, 
     ed. W. H. G. Lewin {\it et al.} 
     (Cambridge University Press: Cambridge), 126

\bibitem[]{}
{\sc Thorne, K. S. \& \.Zytkow, A. N.} 1977, {\it ApJ}, {\bf 212}, 832

\bibitem[]{}
{\sc Tingay, S. J. {\it et al.}} 1995, {\it Nature}, {\bf 374}, 141

\bibitem[]{}
{\sc Tuchman, Y., Mineshige, S., \& Wheeler, J. C.} 1990, {\it ApJ},
	{\bf 359}, 164 

\bibitem[]{}
{\sc van der Klis, M.} 1994, {\it A\&A}, {\bf 283}, 469

\bibitem[]{}
{\sc van der Klis, M.} 1996, private communication 

\bibitem[]{}
{\sc van Paradijs, J.} 1996, {\it ApJ}, in press.

\bibitem[]{}
{\sc van Paradijs, J. \& McClintock, J. E.} 1994, {\it A\&A}, {\bf 290}, 133

\bibitem[]{}
{\sc van Paradijs, J., \& McClintock, J. E.} 1995, in {\it X-Ray Binaries},
     ed. W. H. G. Lewin {\it et al.}
     (Cambridge University Press: Cambridge), 126

\bibitem[Vishniac \& Brandenburg\ 1996]{VB96}
{\sc  Vishniac, E.T., \& Brandenburg, A.} 1996, {\it ApJ}, in press 

\bibitem[Vishniac, Jin, \& Diamond\ 1990]{VJD90}
{\sc Vishniac, E.T., Jin, L., \& Diamond, P.H.} 1990, {\it ApJ}, {\bf 365}, 552

\bibitem[Vishniac \& Diamond\ 1992]{VD92}
{\sc Vishniac, E.T., \& Diamond, P.H.} 1992, {\it ApJ}, {\bf 398}, 561

\bibitem[Vishniac \& Diamond\ 1993]{VD93}
{\sc Vishniac, E.T., \& Diamond, P.H.} 1993, in 
	{\it Accretion Disks in Compact Stellar Systems},
	ed. J.C. Wheeler (World Scientific Press: Singapore), 41

\bibitem[]{}
{\sc Vishniac, E. T. \& Wheeler, J. C.} 1996, {\it ApJ}, in press

\end{thebibliography}
\end{document}